\documentclass[aps,preprint,tightenlines,superscriptaddress,showpacs]{revtex4}
\usepackage{graphicx} 
\usepackage{dcolumn}  

\begin{document}


\preprint{\vbox{ \hbox{   }
                 \hbox{  }
                 \hbox{BELLE-CONF-0768}
         \hbox{  }
}}

\title{ \quad\\[0.5cm]  Measurement of neutral pion pair production\\
in two-photon collisions }


%
\normalsize
\affiliation{Budker Institute of Nuclear Physics, Novosibirsk}
\affiliation{Chiba University, Chiba}
\affiliation{University of Cincinnati, Cincinnati, Ohio 45221}
\affiliation{Department of Physics, Fu Jen Catholic University, Taipei}
\affiliation{Justus-Liebig-Universit\"at Gie\ss{}en, Gie\ss{}en}
\affiliation{The Graduate University for Advanced Studies, Hayama}
\affiliation{Gyeongsang National University, Chinju}
\affiliation{Hanyang University, Seoul}
\affiliation{University of Hawaii, Honolulu, Hawaii 96822}
\affiliation{High Energy Accelerator Research Organization (KEK), Tsukuba}
\affiliation{Hiroshima Institute of Technology, Hiroshima}
\affiliation{University of Illinois at Urbana-Champaign, Urbana, Illinois 61801}
\affiliation{Institute of High Energy Physics, Chinese Academy of Sciences, Beijing}
\affiliation{Institute of High Energy Physics, Vienna}
\affiliation{Institute of High Energy Physics, Protvino}
\affiliation{Institute for Theoretical and Experimental Physics, Moscow}
\affiliation{J. Stefan Institute, Ljubljana}
\affiliation{Kanagawa University, Yokohama}
\affiliation{Korea University, Seoul}
\affiliation{Kyoto University, Kyoto}
\affiliation{Kyungpook National University, Taegu}
\affiliation{\'Ecole Polytechnique F\'ed\'erale de Lausanne (EPFL), Lausanne}
\affiliation{University of Ljubljana, Ljubljana}
\affiliation{University of Maribor, Maribor}
\affiliation{University of Melbourne, School of Physics, Victoria 3010}
\affiliation{Nagoya University, Nagoya}
\affiliation{Nara Women's University, Nara}
\affiliation{National Central University, Chung-li}
\affiliation{National United University, Miao Li}
\affiliation{Department of Physics, National Taiwan University, Taipei}
\affiliation{H. Niewodniczanski Institute of Nuclear Physics, Krakow}
\affiliation{Nippon Dental University, Niigata}
\affiliation{Niigata University, Niigata}
\affiliation{University of Nova Gorica, Nova Gorica}
\affiliation{Osaka City University, Osaka}
\affiliation{Osaka University, Osaka}
\affiliation{Panjab University, Chandigarh}
\affiliation{Peking University, Beijing}
\affiliation{University of Pittsburgh, Pittsburgh, Pennsylvania 15260}
\affiliation{Princeton University, Princeton, New Jersey 08544}
\affiliation{RIKEN BNL Research Center, Upton, New York 11973}
\affiliation{Saga University, Saga}
\affiliation{University of Science and Technology of China, Hefei}
\affiliation{Seoul National University, Seoul}
\affiliation{Shinshu University, Nagano}
\affiliation{Sungkyunkwan University, Suwon}
\affiliation{University of Sydney, Sydney, New South Wales}
\affiliation{Tata Institute of Fundamental Research, Mumbai}
\affiliation{Toho University, Funabashi}
\affiliation{Tohoku Gakuin University, Tagajo}
\affiliation{Tohoku University, Sendai}
\affiliation{Department of Physics, University of Tokyo, Tokyo}
\affiliation{Tokyo Institute of Technology, Tokyo}
\affiliation{Tokyo Metropolitan University, Tokyo}
\affiliation{Tokyo University of Agriculture and Technology, Tokyo}
\affiliation{Toyama National College of Maritime Technology, Toyama}
\affiliation{Virginia Polytechnic Institute and State University, Blacksburg, Virginia 24061}
\affiliation{Yonsei University, Seoul}
  \author{K.~Abe}\affiliation{High Energy Accelerator Research Organization (KEK), Tsukuba} 
  \author{I.~Adachi}\affiliation{High Energy Accelerator Research Organization (KEK), Tsukuba} 
  \author{H.~Aihara}\affiliation{Department of Physics, University of Tokyo, Tokyo} 
  \author{K.~Arinstein}\affiliation{Budker Institute of Nuclear Physics, Novosibirsk} 
  \author{T.~Aso}\affiliation{Toyama National College of Maritime Technology, Toyama} 
  \author{V.~Aulchenko}\affiliation{Budker Institute of Nuclear Physics, Novosibirsk} 
  \author{T.~Aushev}\affiliation{\'Ecole Polytechnique F\'ed\'erale de Lausanne (EPFL), Lausanne}\affiliation{Institute for Theoretical and Experimental Physics, Moscow} 
  \author{T.~Aziz}\affiliation{Tata Institute of Fundamental Research, Mumbai} 
  \author{S.~Bahinipati}\affiliation{University of Cincinnati, Cincinnati, Ohio 45221} 
  \author{A.~M.~Bakich}\affiliation{University of Sydney, Sydney, New South Wales} 
% ...
\collaboration{The Belle Collaboration}

\begin{abstract}
The differential cross section 
of the process $\gamma \gamma \to \pi^0 \pi^0$
has been measured in the kinematical range
0.6~GeV $< W < 4.0$~GeV and $|\cos \theta^*|<0.8$
in energy and pion scattering angle, respectively,
in the $\gamma\gamma$ center-of-mass system. We find at least
four resonant structures including a peak from $f_0(980)$.
In addition, there is evidence for $\chi_{c0}$  production.
We  also make a preliminary discussion of the angular
dependence and cross section ratio to $\gamma \gamma
\to \pi^+\pi^-$.
\end{abstract}

\maketitle
\tighten
\normalsize

{\renewcommand{\thefootnote}{\fnsymbol{footnote}}

\setcounter{footnote}{0}

\normalsize

%
%
%


\section{Introduction}
  Measurements of exclusive hadronic final states in two-photon
collisions provide valuable information concerning the physics of light and 
heavy-quark
resonances, perturbative and non-perturbative QCD 
and hadron-production mechanisms. So far, we have measured
the production cross sections of charged-pion pairs~\cite{mori,nkzw}, 
charged and neutral-kaon pairs~\cite{nkzw,wtchen}, 
and proton-antiproton pairs~\cite{kuo}.
We have also analyzed $D$-meson-pair production to search for a new
charmonium state~\cite{z3930}.

 In this report, we show an analysis of 
neutral-pion pair production in two-photon processes.
The motivation for this study is essentially the same as that
for the charged-pion pair case. But
the two processes are physically different and 
independent; we cannot predict
very precisely what happens in one by measuring only the other.

In the low energy region ($W < 1.0$~GeV), it is expected 
that the difference of meson electric charges 
plays an essential 
role in the difference between the $\pi^+\pi^-$ and 
$\pi^0\pi^0$ cross sections. The predictions are not easy
because of non-perturbative effects.
In the intermediate energy range ($1.0<W<2.4$~GeV), the formation
of meson resonances decaying to $\pi\pi$ is the dominant contribution.  
For ordinary $q\bar{q}$ mesons conserving isospin in 
decays to $\pi\pi$, the only allowed  $I^GJ^{PC}$ states
produced by two photons are  
$0^+$(even)$^{++}$, that is, $f_{J={\rm even}}$ mesons. 
The ratio of the $f$-meson's branching fractions,
${\cal B}(f \to \pi^0\pi^0)/{\cal B}(f \to \pi^+\pi^-)$
is 1/2 from isospin
invariance. But, interference of the resonance 
with the continuum component which cannot be precisely 
calculated distorts the ratio even near the resonant
peaks. The $\pi^0\pi^0$ channel has an advantage in the study of
resonances, since a smaller contribution from the 
continuum is expected in it than in the $\pi^+\pi^-$ channel.

For higher energies, we can invoke a quark model.
In the leading order calculations~\cite{bl,bc}
which take into account
spin correlation between quarks, the $\pi^0\pi^0$
cross section is predicted to be much smaller
than that of $\pi^+\pi^-$, and the cross section ratio 
of  $\pi^0\pi^0$ to $\pi^+\pi^-$ is
around 0.03-0.06. However, higher-order or
non-perturbative QCD effects can modify the
ratio. For example, the handbag model which
considers soft hadron exchange predicts
the same amplitude for the two processes,
and this ratio becomes 0.5~\cite{handbag}. Analyses
of energy and angular distributions of the cross sections
are essential for determining properties of the observed
resonances and for testing the validity of QCD models.

 We present preliminary results of the
measurement of the differential cross section, $d\sigma/d|\cos \theta^*|$,
for the process $\gamma \gamma \to \pi^0 \pi^0$ in
a wide two-photon center-of-mass (c.m.) 
energy ($W$) range from 0.6 to 4.0~GeV,
and a c.m. angular range, $|\cos \theta^*| <0.8$.
(Although we do not need to put the absolute-value symbol
for $\cos \theta^*$ for the present channel where the identical
particle pairs appear in both initial and final states,
we follow the usual convention of the two-photon differential
cross section to avoid unnecessary confusion.)
Our data sample is by several hundred times larger
than in previous experiments~\cite{prev1, prev2}.

\section{Experimental apparatus and trigger}

We use a data sample that corresponds to an integrated luminosity of
95~fb$^{-1}$ recorded with the Belle detector at the KEKB 
asymmetric-energy $e^+e^-$ collider~\cite{kekb}. The energy 
of the accelerator was set at 10.58~GeV (83~fb$^{-1}$),
10.52~GeV(9~fb$^{-1}$), 10.36~GeV($\Upsilon(3S)$ runs,
2.9~fb$^{-1}$) and 10.30~GeV(0.3~fb$^{-1}$), in the c.m.
energy of the $e^+e^-$ beams. 
The difference of the two-photon flux (luminosity function) 
in the measured $W$ regions due to the different beam energies
is very small (a few percent at maximum), and
the fraction of integrated luminosity of the runs with the
lower beam energies is very small. We therefore combine the results for
different beam energies. The effect on the cross 
section deviation is less than 0.5\%.
The analysis is made in the ``zero-tag'' mode, where
neither the recoil electron nor positron is detected. We
restrict the virtuality of the incident photons to be small
by imposing 
strict transverse-momentum balance with respect to the beam axis
for the final-state hadronic system.

  A comprehensive description of the Belle detector is
given elsewhere~\cite{belle}. We mention here only those
detector components which are essential for the present measurement.
Charged tracks are reconstructed from hit information in a central
drift chamber (CDC) located in a uniform 1.5~T solenoidal magnetic field.
The detector solenoid is along the $z$ axis which points
in the direction opposite to that of the positron beam. The CDC and
a silicon vertex detector measure the
longitudinal and transverse momentum components (along the $z$ axis and
in the $r\varphi$ plane, respectively).  Photon detection and
energy measurements are performed with a CsI(Tl) electromagnetic
calorimeter (ECL).

In the present measurement, we require that there be no
reconstructed tracks coming from the vicinity of
the nominal collision point. Therefore, the CDC is used 
for vetoing events with track(s). Photons from 
decays of two neutral pions are detected and their 
momentum vectors are measured by the ECL. 

Signals from the ECL are used to trigger the signal events.
The conditions of the ECL trigger
are as follows: The ECL total energy deposit
in the triggerable acceptance region (see the next subsection)
is greater than 1.15~GeV (the 'HiE' trigger), or,
the number of the ECL clusters counted according
to the energy threshold at 110~MeV for segments
of the ECL is four or greater (the 'Clst4' trigger).
The above energy thresholds are determined from
the experimental data from a study of the correlations
between the two triggers.  No software
filterings are applied for triggering
events by either or both of the two ECL triggers.

\section{Event selection}
The selection conditions for signal candidates of $\gamma \gamma \to
\pi^0 \pi^0$ are as follows:  All the variables 
in the criteria (1)-(5) are measured in the laboratory frame;
(1) there is no good track 
that satisfies $dr < 5$~cm, $|dz| < 5$~cm and $p_t > 0.1$~GeV/$c$,
where $dr$ and $dz$ are the radial and axial distances,
respectively, of the closest approach (as seen in the $r\varphi$ plane) 
to the nominal collision point, and  the $p_t$ is the transverse 
momentum measured in the
laboratory frame with respect to the $z$ axis;
(2) the events are triggered by the HiE or Clst4 triggers;
(3) there are two or more photons whose energy is greater than
100~MeV;
(4) there are just two $\pi^0$'s, each having
transverse momentum greater than 0.15~GeV/$c$,
and each of the decay-product photons has energy greater
than 70~MeV, where the $\pi^0$ is reconstructed from two photons
and is selected with a $\chi^2$ value of a mass constraint fit; 
(5) the total energy deposit in ECL
is smaller than 5.7~GeV.

Then, the transverse momentum in the $e^+e^-$ c.m. frame
($|\Sigma {\bf p}_t^*|$) of the two-pion system
is calculated. For further analysis, (6) we use
events with $|\Sigma {\bf p}_t^*| < 50$~MeV/$c$
as the signal candidates. 

In order to reduce
uncertainty from the efficiency of the hardware ECL triggers, 
we set offline selection criteria which emulate the hardware trigger 
conditions as follows: 
(7) the ECL energy sum within the triggerable region
is greater than 1.25~GeV, {\it or}  all the four photons
composing the two $\pi^0$ are contained in
the triggerable acceptance region. Here, we
define the triggerable acceptance region as 
the polar-angle range in the laboratory system 
$17.0^\circ < \theta < 128.7^\circ$.

\section{Signal candidates and backgrounds}
\subsection{Yield distribution of the signal candidates}
We derive the c.m. energy $W$ of the two-photon collision 
from the invariant mass of the two neutral pion system.
We calculate cosine of the scattering angle of 
$\pi^0$ in the $\gamma \gamma$ c.m. frame, $| \cos \theta^*|$ 
for each event.  We use the $e^+e^-$ collision 
axis in the $e^+e^-$ c.m. frame as the reference of the 
polar angle as an approximation, because
we do not know the exact $\gamma\gamma$ collision axis.

The two-dimensional yield distribution of the 
selected events is shown by the lego plot 
in Fig.~1. The $W$ distribution with $|\cos \theta^*|<0.8$
is shown in Fig.~2(a). We find at least four resonant
structures: clear peaks from $f_0(980)$ near 0.98~GeV and 
$f_2(1270)$ near 1.25~GeV, and broad structures near
1.65~GeV and near 1.95~GeV. Figure 2(b) is an enlarged view of the charmonium
region for  $|\cos \theta^*|<0.4$, where we see some
hints of charmonium contributions in the $\chi_{c0}$($\sim 3.40$~GeV) 
and $\chi_{c2}$($\sim 3.55$~GeV) mass region.

\subsection{Background subtraction}
We study the $p_t$-balance distribution, i.e.,
the event distribution in $|\Sigma {\bf p}_t^*|$
to separate the signal and background components.
The signal Monte Carlo (MC) events show that the signal 
component peaks around 10-20~MeV/$c$ in this
distribution.  In the experimental data, however, in addition to such a
signal component, we find some
contribution from the $p_t$-unbalanced components 
in the low-$W$ region. 
The source of such $p_t$-unbalanced components, in general, is
considered to be non-exclusive processes such as $\pi^0\pi^0\pi^0$ 
etc.  But, in the present case, the background
found in the experimental data is very large only in the low $W$
region where the $\pi^0\pi^0\pi^0$ contribution is expected  
to be much smaller than $\pi^0\pi^0$ in two-photon collisions.
(Note that a $C$-parity odd system cannot go to $\pi^0\pi^0$.) We 
believe that the backgrounds are dominated by beam-background photons (or
neutral pions from secondary interactions) or spurious hits 
in the detector.

Figures 3 (a) and (b) show the $p_t$-balance distributions 
in the low $W$ region. With the fit described below, 
we separate the signal components from the background.
In the intermediate or higher energy regions, 
the backgrounds are less than 1\%, buried under 
the $f_2(1270)$ peak (Fig.~3(c)),
or we do not see any detectable $p_t$-unbalanced components within
the statistical errors (as shown in Fig.~3(d)).  Even at the highest energy
3.6~GeV$<W<4.0$~GeV, $|\cos \theta^*|<0.4$, we find no visible
contamination from the background.

 A fit to the $p_t$-balance distribution is made
to separate the signal and background components for 
the $W$ region below 1.2~GeV.  The fit function is a sum
of the signal component and the background component.
The signal component is taken as an empirical function
reproducing the shape of the signal MC, $y=Ax/(x^{2.1}+B+Cx)$, where
($x \equiv |\Sigma {\bf p}_t^*|$, $A$, $B$ and $C$ are the
fitting parameters, and $y$ is the distribution). This function
has a peak at $x=(\frac{B}{1.1})^{\frac{1}{2.1}}$ and vanishes
at $x=0$ and infinite $x$. The shape of the 
background is taken as a linear function $y=ax$ for $x<0.05$~GeV/$c$ which is 
smoothly connected to a quadratic function above $x>0.05$~GeV/$c$.

The background yields obtained from the fits are fitted to a 
smooth two-dimensional function of ($W$, $|\cos \theta^*|$) 
in order to minimize statistical fluctuations. Then,
the backgrounds are subtracted from the experimental 
yield distribution.  The background yields integrated over the
angle is shown in Fig.~2(a).
We omit the data points in the small-angle ($|\cos \theta^*|>0.6$) 
region in $W<0.72$~GeV, because there the background
dominates the yield.

\begin{figure}
\centering
\includegraphics[width=11cm]{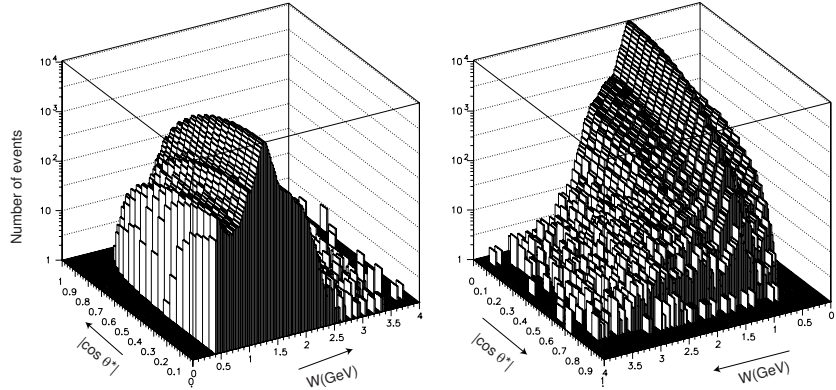}
\label{fig4}
\centering
\caption{The two-dimensional distribution
of the experimental $\pi^0\pi^0$ candidates.
The same distribution is viewed from
two different directions.}
\end{figure}

\begin{figure}
\centering
\includegraphics[width=11cm]{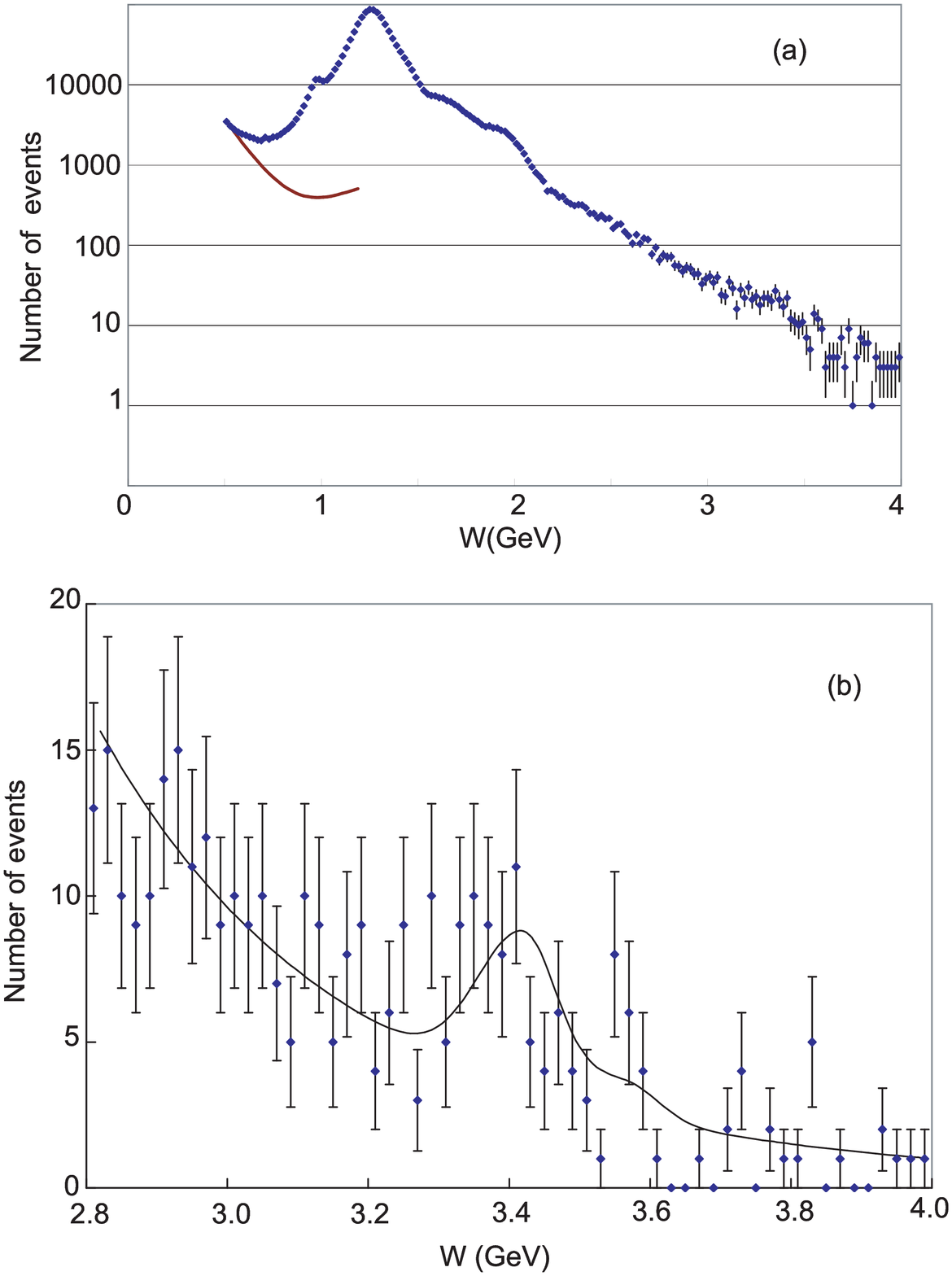}
\label{fig5}
\centering
\caption{The $W$ distribution
of the candidate events. (a) $|\cos \theta^*|<0.8$.
The curve is an estimate of backgrounds from events with
$p_t$ imbalance.
(b)  $|\cos \theta^*|<0.4$, near the charmonium region.
The curve is the fit described in Sect.~VIII.}
\end{figure}

\begin{figure}
\centering
\includegraphics[width=11cm]{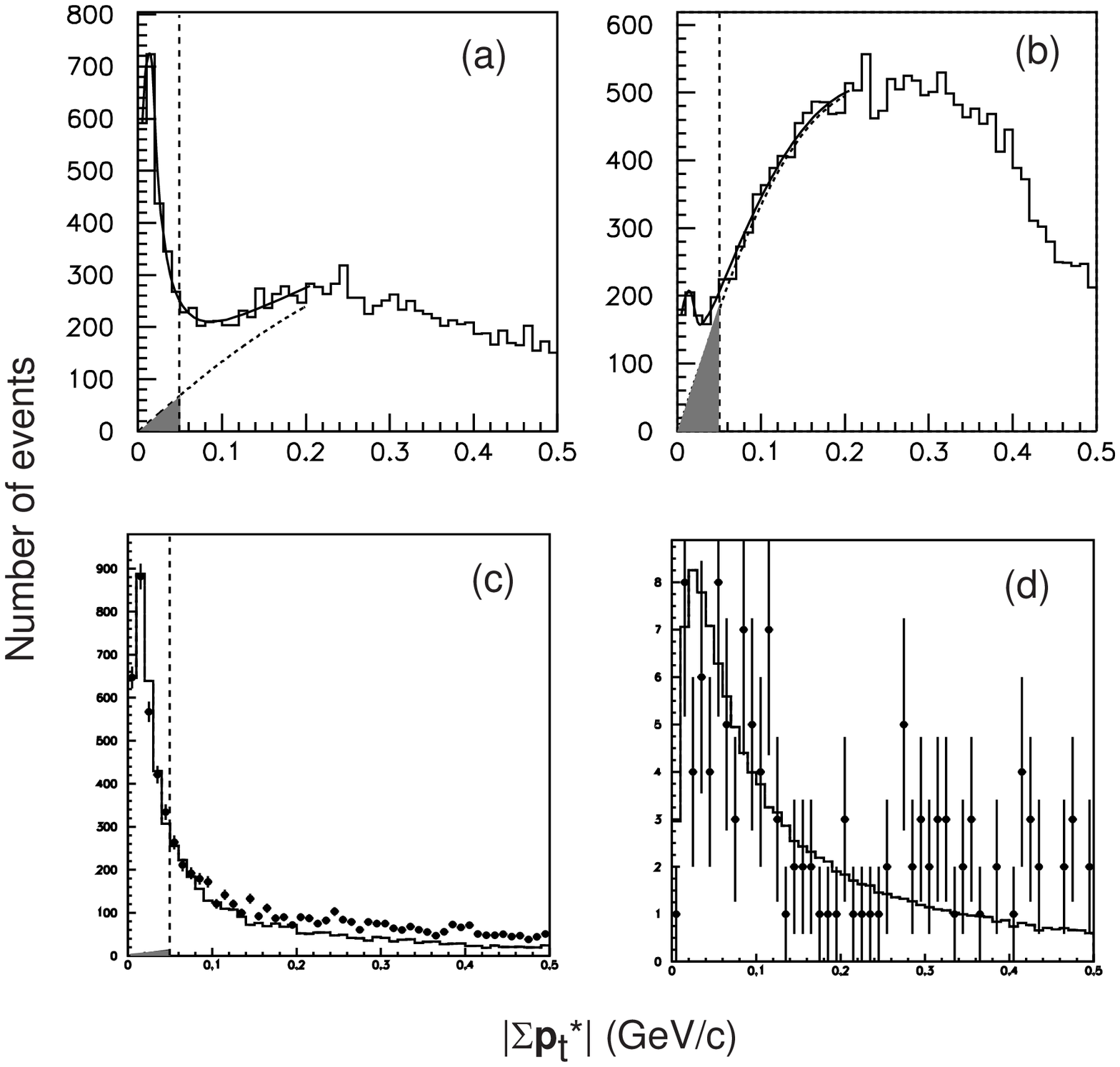}
\label{fig6}
\centering
\caption{The distribution of imbalance in $p_t$
for candidate events. (a) In the bin centered at
$W=0.90~$GeV and $|\cos \theta^*|=0.05$
(The bin width is 0.04~GeV and 0.1 in the $W$ and 
$|\cos \theta^*|$ directions, respectively, in (a)-(c).), 
the experimental distribution (histogram) is fitted with the sum
of signal and background components (curves). The grey
region shows the estimated background contamination in
the signal region. (b) $W=0.66~$GeV, same as in (a) 
for the others. (c)  In the bin of 
$W=1.18$~GeV, $|\cos \theta^*|=0.65$, the experimental
distribution (dots with error bars) is compared with
the signal MC (histogram). The grey region shows the 
estimated background contaminating the signal region
obtained from the fit. (d) For $3.6<W<4.0$~GeV
and  $|\cos \theta^*|<0.4$, the experimental
distribution (dots with error bars) is compared with
the signal MC (histogram).}
\end{figure}

\subsection{Unfolding the $W$ distributions}
We estimate the invariant-mass resolutions from studies 
of the signal-MC and experimental events. 
We find that the MC events show a relative invariant-mass 
resolution of 1.4\%, which is almost constant in the whole $W$ 
region of the present measurement.  
The momentum resolution of $\pi^0$ is known to be 
worse by about 15\% in the experimental data than in MC 
from a study of the $p_t$-balance distributions (described in 
Sect.~V). Moreover, the distribution of the MC is 
asymmetric; it has a longer tail on the lower mass side.

An asymmetric Gaussian function with standard deviations of 
1.9\%$W$ and 1.3\%$W$ on the lower and higher sides of the peak, 
respectively, is used and approximates the smearing reasonably well.  

We calibrate the experimental energy scale and 
invariant-mass distribution using the $\gamma \gamma$
invariant mass from experimental samples 
of $\eta' \to \gamma \gamma$ from two-photon processes.
The peak position is consistent with the nominal mass of
$\eta'$ with an accuracy better than 0.2\%.

 This invariant-mass resolution is comparable to 
or larger than the $W$ bin width (20~MeV) used 
in Figs.~(1) and (2). We made an unfolding of the 
invariant-mass distribution in each $|\cos \theta^*|$ bin
separately, to correct for migrations of signal 
yields between different $W$ bins, 
based on the smearing function of 
the above asymmetric Gaussian shape.
The migration in the  
$|\cos \theta^*|$  direction is expected to be
small and is neglected. 

The unfolding is made using the Singular Value Decomposition (SVD) 
algorithm~\cite{svdunf} in the
yield level, and is applied so as to reproduce 
the corrected $W$ distribution in the 0.9 - 2.4~GeV region,
using data at observed $W$ between 0.72 and 3.0~GeV.
For lower energies, $W < 0.9$~GeV, the effect of the migration is 
expected to be small because of the smaller 
invariant-mass resolution compared
with the bin width. For higher energies, $W > 2.4$~GeV, 
where the statistics is relatively low and
the unfolding would enlarge the errors,
we rebin the result with a bin width of 100~MeV, instead
of unfolding, as described in Sect.~IV.C. Distributions before and after
the unfolding for a typical angular bin ($|\cos \theta^*|=0.225$) 
are shown in Fig.~4.

\begin{figure}
\centering
\includegraphics[width=10cm]{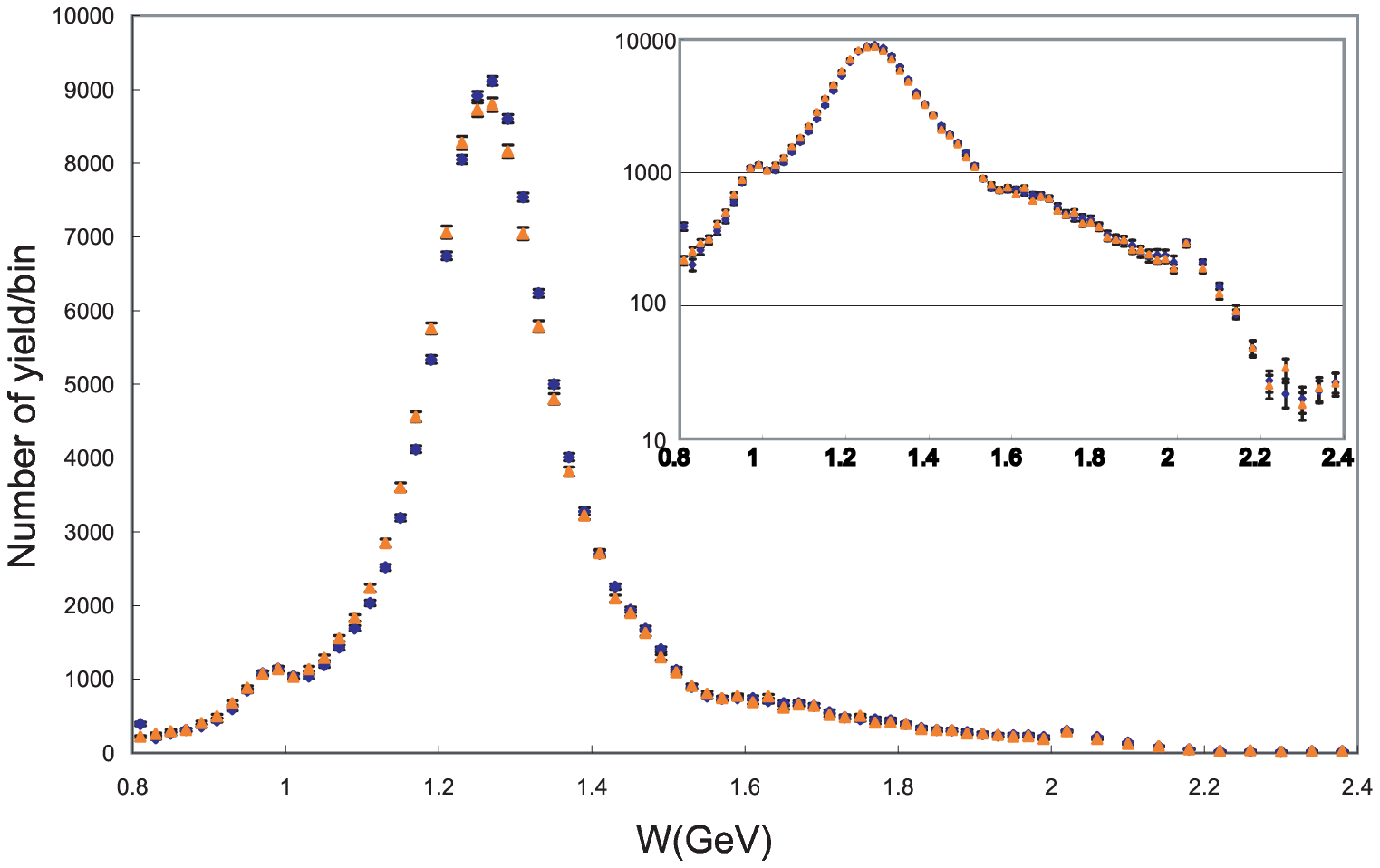}
\label{fig17}
\centering
\caption{Invariant mass distributions of
the yield in each bin before(orange colored
triangles) and
after (dark-blue diamonds) the unfolding,
at $|\cos \theta^*|=0.225$.
The inset shows a logarithmic plot
of the same distribution. The bin width is
different for the regions $W$ below and above 2.0~GeV.}
\end{figure}

\section{Determination of efficiency}

We determine the trigger efficiency for
signal events using the detector and trigger
simulators applied to the signal MC events. 

 The signal MC events for $e^+e^- \to e^+e^- \pi^0\pi^0$ are
generated using TREPS code~\cite{treps} for the trigger efficiency
study at 27 fixed $W$ points between 0.5 and 4.1~GeV,
isotropic in $|\cos \theta^*|$. The angular distribution at 
the generator level does not play a role for the efficiency
determination, because we calculate the efficiencies separately 
in each $|\cos \theta^*|$ bin with width 0.05.
$5 \times 10^4$ events are generated at each $W$ point.

To minimize statistical fluctuation in the MC, we fit
the numerical results of the trigger efficiency to a
two-dimensional empirical function in $(W, |\cos \theta^*|)$.
The $W$ dependences are shown in Fig.~5 for two typical
angular bins. We find that the trigger efficiency is
almost flat and close to 100\% for the region $W$ 
above 1.3~GeV. The angular
dependence is rather small.

 Separately, we generated signal
MC events at 48 $W$ points in the same $W$ region
with $10^5$ events at each value of $W$, for the acceptance
calculations. Here we call the efficiency of the
selections not including the hardware trigger 
``acceptance''; the net efficiency is a product of
the trigger efficiency and the acceptance. 
The determined acceptance from the MC
events is fitted by a smooth
two-dimensional function of ($W$, $|\cos \theta^*|$)
(Fig.~6). It is about 11\% at maximum and gets smaller
(down to around 1\%) at lower $W$ or smaller c.m.
angle (larger $|\cos \theta^*|$).

The acceptance calculated using the signal MC events
is corrected for a systematic difference found
between the peak widths in the $p_t$-balance
distributions of the experimental data and
the MC, which could affect the acceptance
through the $|\Sigma {\bf p}_t^*|$ cut. 
It originates from a difference
in the momentum resolution for $\pi^0$ between
data and MC.
We find that the peak position of the experimental
data in the distribution is 10\% to 20\% higher than the MC
expectation, depending on $W$ and $|\cos \theta^*|$. 
The correction factor ranges from 0.90 to 0.95.


\begin{figure}
\centering
\includegraphics[width=8.5cm]{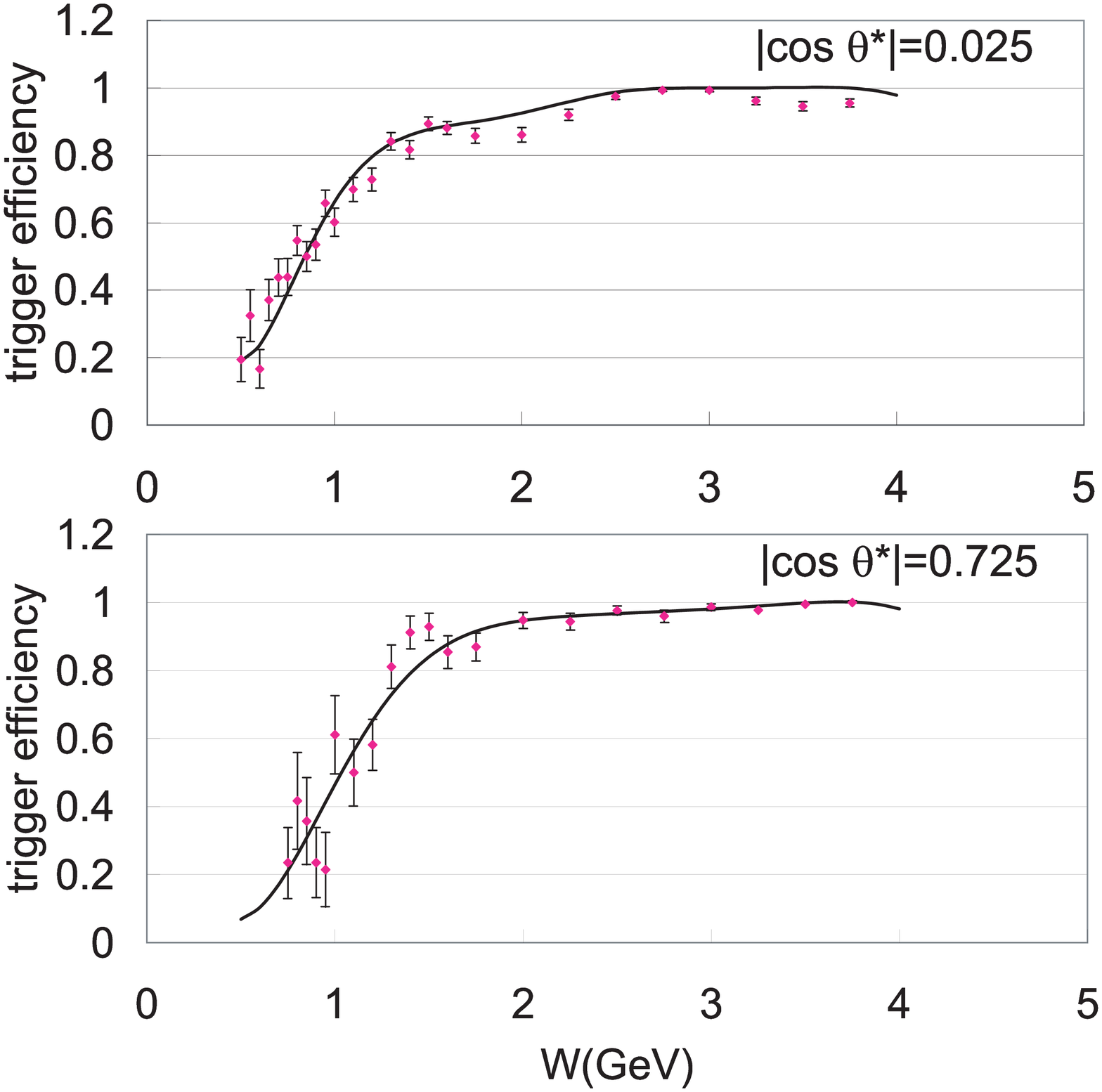}
\label{fig3}
\centering
\caption{The $W$ dependence of the
trigger efficiency for two angular bins.
The curves are the fit to parameterize it.}
\end{figure}

\begin{figure}
\centering
\includegraphics[width=8cm]{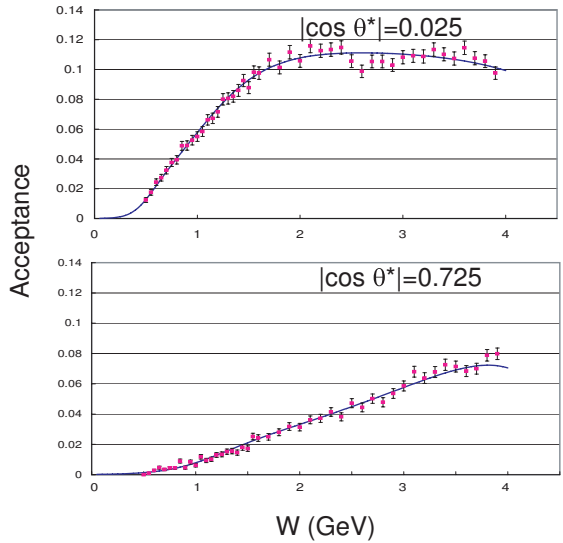}
\label{fig8}
\centering
\caption{The $W$ dependence of the
acceptance for two angular bins.
The curves are the fit to parameterize it.
No correction for the $\pi^0$ momentum resolution
is included.}
\end{figure}

\section{Cross section calculation}

The differential cross section for each
($W$, $|\cos \theta^*|$) point is derived
by the following formula:

\[
\frac{d\sigma}{d|\cos \theta^*|} =
\frac{\Delta Y - \Delta B}{\Delta W \Delta |\cos \theta^*| 
\int{\cal L}dt L_{\gamma\gamma}(W)  \eta_{\rm trg} \eta_{\rm acc} }
\]
where $\Delta Y$ and $\Delta B$ are the signal yield and
the estimated $p_t$-unbalanced background in the bin, 
$\Delta W$ and $\Delta |\cos \theta^*|$ are the bin widths, 
$\int{\cal L}dt$ and  $L_{\gamma\gamma}(W)$ are
the integrated luminosity and two-photon luminosity function
calculated by TREPS, respectively, and  $\eta_{\rm trg}$ and
$\eta_{\rm acc}$ are the trigger efficiency and the acceptance, respectively,
the latter including the correction described in the previous section.

Figures~7(a) and (b) show the $W$ dependence of the
cross section integrated over $|\cos \theta^*| < 0.8$ 
and $|\cos \theta^*| < 0.6$, respectively. 
They are obtained by simply adding
$d\sigma/d|\cos \theta^*| \cdot  \Delta |\cos \theta^*|$ 
over the corresponding angular bins.

The data points for 0.9~GeV $ < W <$ 2.4~GeV are
the unfolded results with the bin widths $\Delta W=0.02$~GeV
(0.04~GeV) for $W$ above (below) 2.0~GeV.
For the data points above 2.4~GeV, we average five data 
points each with a bin width of 0.02~GeV and get results
for every 0.1~GeV at $W$. We have removed the bins in
the range 3.3~GeV $< W < 3.6$~GeV, because we cannot separate 
the components of the $\chi_{c0}$, $\chi_{c2}$  
and the continuum in a model-independent way due to a finite 
mass resolution and insufficient statistics in the measurement.

We show the angular dependence
of the differential cross section
at several $W$ points in Fig.~8.

It is noted that the cross section results
after the unfolding are no longer
independent of each other in the neighboring bins,
in both central values and size of errors.

\begin{figure}
\centering
\includegraphics[width=14cm]{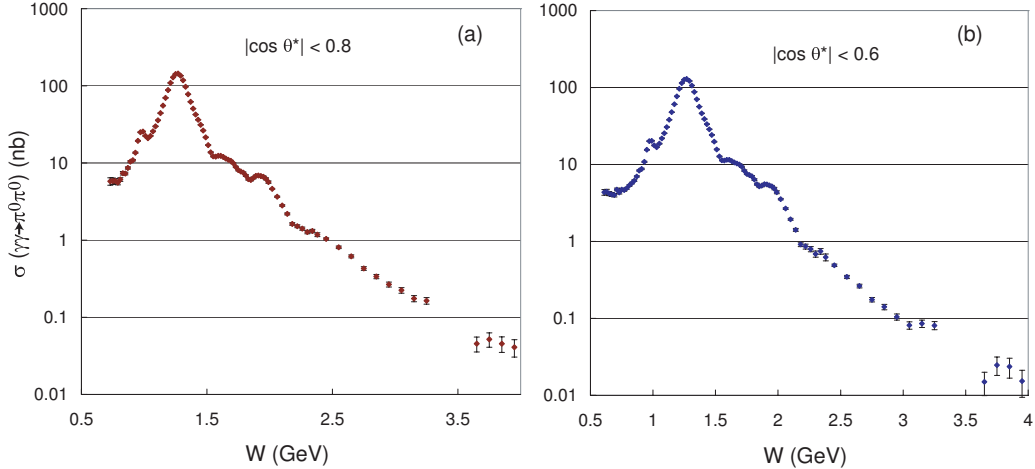}
\label{fig18}
\centering
\caption{The cross section results
integrated in the angular regions
(a) $|\cos \theta^*|<0.8$ and
(b) $|\cos \theta^*|<0.6$ after
the unfolding and rebinning.
}
\end{figure}

\begin{figure}
\centering
\includegraphics[width=11cm]{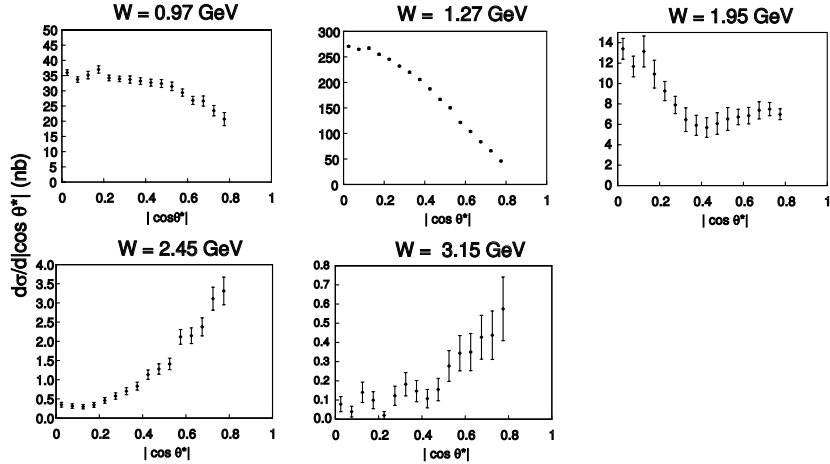}
\label{fig20}
\centering
\caption{The differential cross sections
for five selected $W$ points, 0.97~GeV,
1.27~GeV, 1.95~GeV, 2.45~GeV and 3.15~GeV.
The results of the first three $W$ points
are after the unfolding. }
\end{figure}

\section{Systematic errors}
We summarize the evaluation of the systematic errors
for $\sigma(|\cos \theta^*| <0.8)$ ($\sigma(|\cos \theta^*| <0.6)$ 
with $W<0.72$~GeV) for 
each $W$ point. They come from the following major error sources:\\
{\it Trigger efficiency}:  The systematic error of 
the Clst4 trigger is assigned as $2/3$ of the difference
of the efficiencies with the different threshold
assumptions for the ECL cluster --
110~MeV and 100~MeV -- set in the trigger simulator for the
energy region $W<2.5$~GeV.
Separately, we take the uncertainty in the HiE trigger
efficiency as 4\% for the whole $W$ region. The systematic
errors from the two triggers are added in quadrature.
This systematic error is large in the low $W$ region, 20\%-30\% 
for $W<0.8$~GeV.\\
{\it The reconstruction efficiency}: 6\% for two pions.\\
{\it The  $p_t$-balance cut}: 3\%-5\%. One half of the 
correction size discussed in Sect.~V.\\
{\it Background subtraction}: 20\% of the size of the 
subtracted component is assigned as the error from this source.
In the $W$ region where the background subtraction 
is not applied ($W>1.2$~GeV), we neglect the error for 1.2~GeV$<W<1.5$~GeV,
or assign 3\% for $W>1.5$~GeV, considering an upper limit 
of the background contamination expected from the $p_t$-balance
distributions.} \\
{\it Luminosity function} 4-5\%. 4\% (5\%) for $W<(>) 3.0$~GeV.

 The total systematic error is 9\% in the intermediate $W$
region, 1.05~GeV$<W<2.7$~GeV. It is much larger for lower
$W$, 15\% at $W=0.85$~GeV, 30\% at $W=0.70$~GeV
and 55\% at $W=0.61$~GeV. The error is dominated
by the background subtraction for low $W$.
For higher $W$, the systematic error is rather
stable, 10\%-11\% for 2.7~GeV$<W<4.0$~GeV.

\section{Discussion}
 We compare our results with the previous measurements by
Crystal Ball at DORIS II~\cite{prev2} (Fig.~9). The 
agreement is fairly good. The error bars from the
two experiments are statistical only, and
the systematic errors ( 7\% (11\%) for $W>(<)~0.8$~GeV 
for the Crystal Ball results) should also be considered
in the comparison. The present measurement
has several hundred times more statistics than the Crystal Ball
measurement.

\begin{figure}
\centering
\includegraphics[width=9cm]{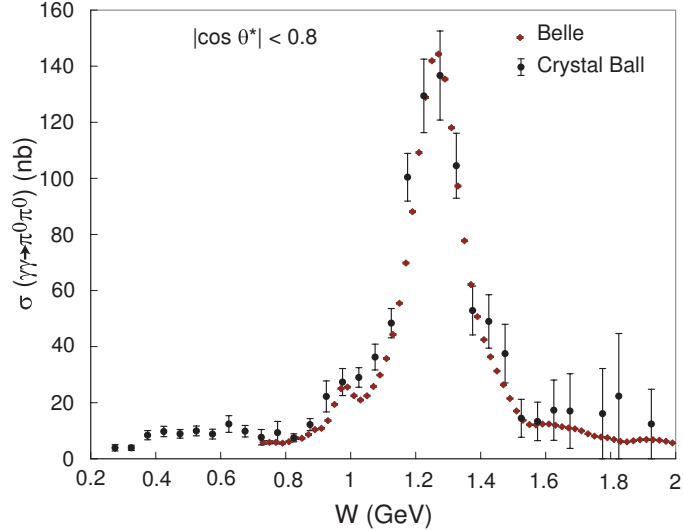}
\label{fig18}
\centering
\caption{The cross section results
integrated over the angular regions
$|\cos \theta^*|<0.8$  
compared with the previous measurement from the 
Crystal Ball~\cite{prev2}.}
\end{figure}

We find prominent resonant structures near 0.98~GeV and
1.27~GeV. They are from $f_0(980)$ and $f_2(1270)$, respectively and  
are observed also in the $\gamma\gamma \to \pi^+\pi^-$ process~\cite{mori}. 
It is for the first time that the $f_0(980)$ is observed
as a clear peak in $\gamma\gamma \to \pi^0\pi^0$.
We find clear evidence for rather broad peaks at 1.65~GeV and 1.95~GeV.
For these, any quick assignment to well known
states is not easy. In addition, 
we find a hint of a possible inflection point near
1.4~GeV and a structure in the mass region of 
two charmonium states $\chi_{c0}$ and $\chi_{c2}$. 
The structures above found 
in the 1.2 - 2.1~GeV region are somewhat similar to the distribution
observed in the $\pi^0\pi^0$ spectrum from the 
$\pi^- p \to \pi^0 \pi^0 n$ experiment, GAMS~\cite{gams}.

 We fit the yield distribution 
in the range 2.8~GeV$<W<4.0$~GeV
and $|\cos \theta^*|<0.4$ including
the contribution from the $\chi_{c0}$ and $\chi_{c2}$
charmonia with a binned maximum-likelihood method.
The fit is shown in Fig.~2(b).
We take into account the finite invariant-mass
resolution effect introduced in the 
last section in the fit. We have fixed
the nominal masses and the width of $\chi_{c0}$
to the world averages~\cite{pdg2006} (we neglect
the width of $\chi_{c2}$).  The background
component is assumed to have the shape of $\sim W^{-n}$. 

 From the fit, we find the yields  $35.3 \pm 9.2$ events
and $8.2 \pm 6.4$ events for $\chi_{c0}$
and $\chi_{c2}$, respectively. These yields
provide the products of the two-photon decay
widths and the branching fractions, 
$\Gamma_{\gamma\gamma}(\chi_{cJ}){\cal B}(\chi_{cJ} \to
\pi^0\pi^0) = 8.4 \pm 2.2 (stat.) \pm 0.8 (syst.)$~eV and
$0.29 \pm 0.23 \pm 0.03$~eV, for $\chi_{c0}$ and $\chi_{c2}$,
respectively. The former and latter provide  
evidence for the $\gamma \gamma \to \chi_{c0} \to \pi^0\pi^0$
signal at the $4.5\sigma$ level and
an upper limit at the
90\% C.L., $ \Gamma_{\gamma\gamma}(\chi_{c2}){\cal B}(\chi_{c2} \to
\pi^0\pi^0) < 0.75$~eV which is obtained from the yield where the
two times the log-likelihood of the fit is smaller by $(1.64)^2$ 
than that of the best fit, respectively.
The above central values show good agreement with 
the world averages of 
$\Gamma_{\gamma\gamma}(\chi_{cJ}){\cal B}$
measured in the $\chi_{cJ} \to \pi^+\pi^-$ process~\cite{pdg2006}, 
considering isospin invariance 
(apply $\pi^0\pi^0:\pi\pi = 1:3$ to the branching
fractions), which includes the Belle measurements
of the process $\gamma\gamma \to \chi_{cJ} \to \pi^+\pi^-$~\cite{nkzw}. 

The general trend of the angular dependence of
the differential cross section is as follows:
The differential cross section has a maximum
at $|\cos \theta^*|=0 $ for the $W<2.1$~GeV region. For the $W>1.9$~GeV,
however, the angular dependence shows a  rise toward the forward
angles. In the higher $W$ region, the point in c.m. angle 
at which the rise in the differential cross section begins 
moves toward the forward direction, 
and the rise gets steeper, as $W$ increases.

\begin{figure}
\centering
\includegraphics[width=10cm]{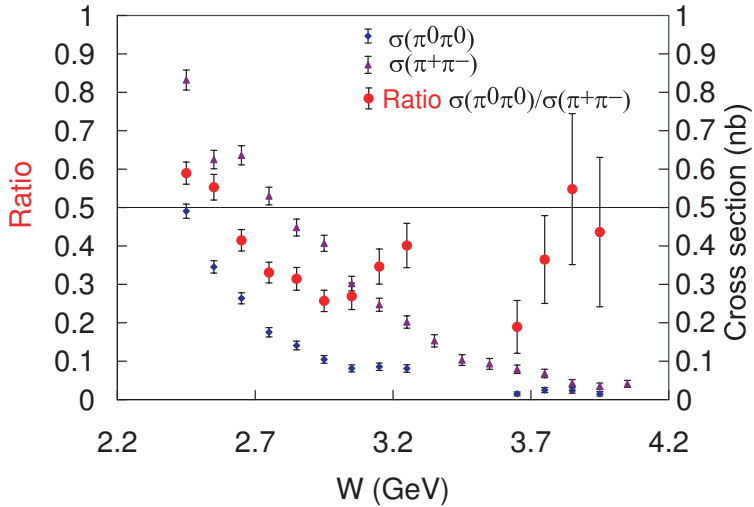}
\label{fig21}
\centering
\caption{The cross sections of the
$\gamma\gamma \to \pi^0\pi^0$ and 
$\gamma \gamma \to \pi^+\pi^-$ reactions
for $|\cos \theta^*|<0.6$. The blue closed
circle and the violet triangles are
the cross sections of the two reactions
in units of nb. The red closed circle
is the ratio. The error bars are
statistical only.}

\end{figure}

We show the cross section ratio between $\gamma \gamma \to \pi^0\pi^0$
and  $\gamma \gamma \to \pi^+\pi^-$~\cite{nkzw} for $|\cos \theta^*|<0.6$ and 
2.4$<W<4.0$~GeV, in Fig.~10. The error bars in the figure are
statistical only. 
Each measurement of the cross section has
typically a 10\% systematic error, partially correlated
with each other for different points.
Several $\gamma \gamma \to \pi^+\pi^-$ measurements
above the charmonium masses have larger systematic
errors, $\sim$ 25\%.

 The energy dependence of the ratio 
above $W >2.7$~GeV is smooth with a value around 
0.3-0.4. This ratio is
larger than the prediction from the lowest-order QCD
calculation. We, however, need more detailed investigations 
of the $W$ and angular dependences in comparison with
the predictions of the QCD models to test the expected
asymptotic nature.

\section{Conclusion}
We have measured the cross section of the process
$\gamma \gamma \to \pi^0\pi^0$ in the $\gamma \gamma$
c.m. energy and angular regions of
0.60~GeV $< W < 4.0$~GeV and $|\cos \theta^*|<0.8$.
In the cross section, several resonant structures 
are seen, including a 
statistically significant peak from the $f_0(980)$. We find
that the angular dependence 
(forward- and/or large-angle enhancements) changes drastically at around 
$W= 2.0$~GeV. The ratio of cross sections of $\gamma \gamma \to
\pi^0\pi^0$ and $\gamma \gamma \to
\pi^+\pi^-$ in the 3 GeV region is also obtained.     
\ \\
\ \\
We thank the KEKB group for the excellent operation of the
accelerator, the KEK cryogenics group for the efficient
operation of the solenoid, and the KEK computer group and
the National Institute of Informatics for valuable computing
and Super-SINET network support. We acknowledge support from
the Ministry of Education, Culture, Sports, Science, and
Technology of Japan and the Japan Society for the Promotion
of Science; the Australian Research Council and the
Australian Department of Education, Science and Training;
the National Natural Science Foundation of China under
contract No.~10575109 and 10775142; the Department of
Science and Technology of India; 
the BK21 program of the Ministry of Education of Korea, 
the CHEP SRC program and Basic Research program 
(grant No.~R01-2005-000-10089-0) of the Korea Science and
Engineering Foundation, and the Pure Basic Research Group 
program of the Korea Research Foundation; 
the Polish State Committee for Scientific Research; 
the Ministry of Education and Science of the Russian
Federation and the Russian Federal Agency for Atomic Energy;
the Slovenian Research Agency;  the Swiss
National Science Foundation; the National Science Council
and the Ministry of Education of Taiwan; and the U.S.\
Department of Energy.

\end{document}